\documentstyle[prl,aps]{revtex}
\input psfig
\begin{document}
\draft
\title{Spin structure of impurity band of semiconductors in two and three-dimensional cases}
\author{I. V. Ponomarev$^{*}$, V. V. Flambaum}
\address{School of Physics, The University of New South Wales,
Sydney 2052, Australia}
\author{A. L. Efros}
\address{Department of Physics, University of Utah, Salt Lake City UT 84112}
\date{\today}
\maketitle
\begin{abstract}
The exchange interaction between  electrons located at different randomly 
distributed impurities is studied for small density of impurities. 
The singlet-triplet splitting  $2J(R)$ is calculated for two Coulomb centers 
at a distance $R$. Interpolated formulas are found which work for all
distances $R$ from zero to infinity. The data from atomic physics are used 
for the interpolation in three-dimensional case. For two-dimensional case the
original calculations are performed to find asymptotic behavior of the 
splitting at large $R$,  the splitting for the ``two-dimensional helium atom''
($R=0$) and the splitting  at $R=a_B$, where $a_B$ is the effective Bohr 
radius. The spin structure of impurity band is described  by the Heisenberg 
Hamiltonian. The ground state of a system consists of localized singlets. 
The  new results are obtained for the distribution  of the singlet pairs in
the ground state. These results are  exact at low density. The problem is 
reduced to a non-trivial geometric problem which is solved in the mean field
approximation and by computer modeling. The density of free electrons is 
found as a function of temperature and the distribution function of the
singlet-triplet transitions energies is calculated. Both functions are given
in an analytical form.
\end{abstract}

\pacs{PACS: 72.15.Cz, 75.10.-b, 31.15.-p}
\section{Introduction}
The structure of the impurity band of semiconductors  has been widely studied 
during the last two decades both theoretically and experimentally 
(See monograph\cite{mon} ). In the early theoretical  studies the spin
structure of the impurity band was completely ignored. Recent experiments 
suggest  that the spin structure is very important for the variable range
hopping conductivity, especially near the metal-non-metal 
transition\cite{shlim,sar,sim,ish,shl,krav}.
 
The exchange interaction must be  the main mechanism of the spin-spin
interaction in the  impurity band. 
It appears as a result of the overlap 
of the wave  functions of different states. The scale of this interaction 
decreases exponentially with increasing distance between the states. Thus, 
this interaction becomes the most important one  near the metal-non-metal 
transition. In this region the scale of the interaction is of the order of the
binding energy of a single impurity.

In this paper we study the spin structure of the impurity band  created by  
Coulomb impurities  in both two and three dimensional cases  in the limit of low
density of impurities.
In the two-dimensional case the impurities  may be located
either outside or inside the plane of  electron gas. We assume that all 
impurities are occupied by one electron. In this case we can consider the
coordinates of the occupied centers as random variables without any 
correlations. 

 Our study of the  spin structure is based upon the Heisenberg Hamiltonian 
which takes into account the spin-spin interaction of the electrons localized
at different randomly distributed impurities.
\begin{equation}
H=\sum_{i\neq k}J_{ik}(I/2+{\bf s_i\cdot s_k}),
\label{ham}
\end{equation}
where ${\bf s}$ is spin 1/2 operator, $I$ is a unit matrix, and $i,k$ denote 
different impurity atoms. The sum is over all pairs of impurities. The density of impurities 
is assumed to be small.

This problem has a long story\cite{rosso,bhattc,lee,andres,bhatt,thomas,fisher,hirsch}.
The following important results have been obtained:\\
1. The ground state of the system consists of local singlets.\\
2. Rosso\cite{rosso}, Thomas and Rosso\cite{thomas}, Andres {\it et al.}\cite{andres} used 
different selfoconsistent approaches to get  the distribution of the 
excitation energies of the  singlet-triplet transitions.\\
3. Bhatt and Lee\cite{bhatt,lee} worked out a computational scaling approach which is exact
at small density of impurities. They have also mentioned a drastic difference
 between the  Heisenberg and 
Ising models.\\
4. As far as we know all previous authors used  simplified versions for the function  $J(R)$.

Our paper pursues the following goals:\\
 1.  We analyze the existing methods  to find the
distribution of excitation energies and propose a new  modification  for
 one-, two- and three-dimensional cases.
 Our approach is exact at low densities and it  allows   to get an
 approximate analytical expression for this distribution.\\
2. 
To get an estimate for  the energy of spin ordering one  needs a reliable 
calculation of the coefficients $J_{ik}$ which are defined here as
$1/2$ of the singlet-triplet splitting for the two 
states corresponding to the impurities  $i$ and $k$.  We have  performed these
calculations  for a pair of the Coulomb centers  at a distance
$R_{ik}$. The result of the computations is a  function $J(R)$ which is 
reliable at all distances from zero to infinity.

The paper is organized as follows.
In Section 2 we consider Hamiltonian Eq.(\ref{ham}) in the case of small
impurity density. We show that the ground state mostly
consists of independent singlets. We show that the problem of finding these 
singlets can be reduced to  a non-trivial geometric problem. We solve it in 
a mean field approximation and by computer modeling. The solution of this 
problem gives the distribution function $F(E)$ for the energies of the 
singlet-triplet transitions for a given function $J(R)$.

 In Section 3 we calculate $J(R)$ and its inverse function $R(J)$. 
 For the $3D$-case we present  interpolated formulae which are based upon  the
results of well-known   calculations  for two hydrogen atoms. These calculations 
include analytical results for  large distances\cite{Gor}, 
numerical calculations at intermediate distances, and known results for the
 singlet-triplet splitting
of the  He atom.  Similar interpolated formulae are presented  for the $2D$-case. 
They are based upon our  original calculations given  in the appendices. 
We present an analytical expression for $J(R)$
at large distances, a numerical result for $J(a_B)$,
and variational calculations for a ``two-dimensional He atom''.

In the Conclusion we discuss the distribution  function of singlet-triplet 
splittings $F(\ln\varepsilon)$, where $\varepsilon=J/J(0)$, 
and the density of free spins $\rho (T)$ at finite 
temperature $T$. These two functions  are the final results of our paper. 

\section{Ground state and excited states of the Heisenberg Hamiltonian in the 
 impurity band}
\subsection{The structure of the ground state}

We  find the ground state and excited states of the Hamiltonian Eq.(\ref{ham})
using the following properties of $J_{ik}$.
\begin{itemize}
\item All  $J_{ik}>0$, which means an antiferromagnetic interaction.  
\item The density of impurities $n$ is assumed to be small, so that the
average distance between them is larger than the characteristic length of the
exponential decay of $J_{ik}$. This means 
there is
 a very large dispersion of $J_{ik}$.
In fact we shall assume that if
$J_{ik}>J_{lm}$, then $J_{ik}\gg J_{lm}$. Thus, we ignore the cases when the
distance $R_{ik}$ is very close to the distance  $R_{lm}$,  assuming that these two pairs
are not very far from each other. 
\end{itemize}
To understand the physics of the problem it is very helpful to consider the
Hamiltonian (\ref{ham}) with four impurities only (Fig. 1a). From a general
principle one can conclude\cite{lan} that the energy spectrum consists of 
six levels, one level with spin $S=2$, three levels with $S=1$, and two levels
with $S=0$. Let us assume that $J_{12}$ is much larger than  all other 
$J_{ik}$ in this problem. Then the ground state wave function 
describes two singlets at sites (1,2) and (3,4). It is easy to write the 
energy of the ground state and the first excited state assuming 
\begin{equation}
J^{\prime}=\sum^{\prime}J_{ik}\ll J_{12},
\label{in}
\end{equation}
where the sum includes all   $J_{ik}$ except $J_{12}$ and $J_{34}$.
The ground state energy $E_0$ and the energy of the first excited state $E_1$ are given by
the equations
\begin{equation}
E_0=-J_{12}-J_{34}+J^{\prime}/2;\quad E_1=-J_{12}+J_{34}+J^{\prime}/2.
\label{en}
\end{equation}
The physical meaning of Eq. (\ref{en}) is simple. Two singlets (1,2)
 and (3,4) do not
interact with each other if condition (\ref{in}) is fulfilled. The $J^{\prime}/2$ 
terms come from the first term in the Hamiltonian (\ref{ham}).

In this approximation  the excitation energy is $E=2J_{34}$. The ground state
has a total spin $S=0$ while the first excited state has $S=1$.

Bhatt and Lee\cite{bhatt,lee} take into account the next approximation for the
excitation energy
\begin{equation}
E=2J_{34}+(J_{13}-J_{23})(J_{24}-J_{14})/J_{12}.
\label{bh}
\end{equation}
Since $J_{12}$ is the largest term, the second term should be small. It looks
like it can change the ground state from singlet to triplet if $J_{34}$ is 
unusually small. However, such configurations are extremely rare. It happens 
because in the case of small $J_{34}$ one should consider Fig.1b with 6 spins
rather than  Fig.1a. Indeed, very small $J_{34}$ means a long distance between
impurities 3 and 4. 
It is more likely that in this situation some other  strong singlet (5,6) is
the  nearest neighbor of the  impurity 4 rather than the singlet (1,2).

 In this 6-spin system we have 2 strongly connected groups
of spins, namely 1,2,3 and 4,5,6. Assume that $J_{12}$ and  $J_{56}$  provide
the strongest bonds in each group. Suppose there is no interaction between the 
groups. Then, the ground state in each of them is a degenerate 
doublet. Altogether the system is 4-fold degenerate. If one takes into 
account $J_{34}$ the degeneracy of the ground state will be lifted. One gets
a singlet and a triplet with the energy splitting  $2J_{34}$. On the other hand,   
the general  6-spin problem can be solved assuming that both $J_{12}$ and 
$J_{56}$ are infinite.  In this approximation one gets the same result: 
the ground state is a singlet and the excitation energy $E=2J_{34}$. It 
follows that the other bonds connecting the two groups, like $J_{35}$ may 
contribute to the excitation energy only in the second order of  perturbation 
theory. This contribution will contain a small dimensionless coefficient like
$J_{35}/J_{12}$ and it may be neglected. Thus,  it is
not necessary to  take into account the renormalization of the weak bonds due to 
their strong neighbors in the limit of small density.  Bhatt and Lee also 
mention\cite{lee} that their 
computations show the  triplet ground state  in very rare cases.

Thus, we assume  that the ground state energy of any even number
of impurities has $S=0$ and the system can be split into localized  singlets.
To find the pairs of impurities which form the singlet in the ground state we 
propose the following geometric problem.\\
1. For every impurity in the system find its nearest neighbor.\\
2. Take the pair with the smallest distance. Generally, the nearest neighbor 
   of a site A does not have site A  as its nearest neighbor. But for the 
   closest pair this is the case.\\
3. This closest pair forms a singlet with the largest binding energy. To find 
   all other singlets remove both sites of the first pair. Go 
   to point 1 and continue until all the singlets will be found.\\
The same geometric problem has been proposed by Thomas and Rosso\cite{thomas} for
three-dimensional case.

 Assuming that all neighboring $J_{ik}$ are very different one can write
the  total energy of the lowest state in the form
\begin{equation}
E_0=-\sum_s J_{ik}+{1\over 2}\sum_{other} J_{ik},
\label{gr}
\end{equation}
where the first sum includes all pairs which form singlets and the second one 
includes all other pairs.

One can prove that the distribution of singlets, obtained as a solution of the
problem above, gives  the minimum of total energy. Suppose, for 
example, that the solution prescribes the  configuration of singlets 
(1,2), (3,4) and (5,6), for impurities with the numbers from 1 to 6. One can 
show that any other location of singlets at the same impurities, like (1,3),
 (2,5) and (4,6), has larger energy.

 We mention first that the contribution to the energy  from all other
impurities like 7,8... is the same at all configurations of singlets of six 
chosen impurities.  Suppose now that $J_{12}\gg J_{34},J_{56}$.  Then
 all other $J_{ik}$ connecting the six impurities
are also  less than $J_{12}$. Indeed, if one of them were larger, it would be used to 
form a singlet instead of $J_{12}$. Thus, any rearrangement of the pairs 
within 6 impurities that destroys singlet (1,2) increases the total energy.
In the same way one can show that rearrangement of singlets in the system of
four impurities 3,4,5,6 also increases the total energy. The same  
consideration can be done for any even number of impurities. Thus, the solution 
of the above geometric problem gives the ground state of the system.

\subsection{Solution of  geometric problem and  distribution function of  excitation
 energies}

We start with the simplest  mean field approximation.
Suppose we are at the stage where all pairs with distance less than $R$ are 
removed and we want to find the residual impurity density $n(R)$. The crucial point of the mean
field approximation  is that we neglect correlations in the positions of the remaining impurities 
except that they  cannot be closer to one another than $R$.

We start with the two-dimensional case. Let us draw a circle around each 
impurity  with the radius $R$. There will be no other impurities  inside the 
circles.  Now increase the radii from $R$ to $R+dR$ and calculate how many 
impurities occur in the rings between $R$ and
$R+dR$. The total number of these impurities  gives the  decrease of $N(R)$, where
$N(R)=S n(R)$ and $S$ is the total area of the system. 
Thus, one gets the equation
\begin{equation}
dN(R)=-N(R)2\pi R  \tilde{n}(R)dR.
\end{equation}
Here $\tilde{n}(R)$ is the density of the impurities outside the circles.
It is slightly larger than $n(R)$ (see below), but in the simplest mean field  approximation 
we ignore this difference.

It is convenient to introduce the dimensionless coordinate $X=\sqrt{\pi n_0}R$  
and the normalized number of particles (or density) $\rho(X)=n/n_0\equiv N/N_0$.
Here $n_0=n(0)$ is the initial concentration of particles. The differential
equation for $\rho(X)$ at $\tilde{n}=n$ has the form

\begin{equation}
d\rho=-2X \rho^2dX.
\end{equation}
The solution of this equation with the condition $\rho(0)=1$  is
\begin{equation}
\rho_2(X)={1\over {1+X^2}},
\label{n}
\end{equation}
where $\rho_2(X)$ is the two-dimensional density.
Similar calculations for the three- and one-dimensional cases give
\begin{equation}
\rho_d(X)={1\over {1+X^d}}.
\label{n3}
\end{equation}
Here 
$X=(4\pi/3n_0)^{1/3}R$  at $d=3$ and  $X=2n_0 R$ at $d=1$. This distribution has been obtained
by Rosso\cite{rosso} for $d=3$. One can show that at small $X$ the above results are exact,
including
$X^d$-corrections. Bhatt\cite{bhattc} has pointed out that it is not exact at large $X$. 
We believe that the exact distribution has a following form at large $X$ 
\begin{equation}
\rho_d(X)={1\over b_d X^d},
\label{b}
\end{equation}
where  the coefficient $b_d \neq 1$  and it  depends on the  dimensionality of space $d$.
It follows from Eq.(\ref{b}) that the average density $n(R)$ is 
independent of $n_0$ at large values of $R$
and it is of the order of $R^{-d}$. This is because the average distance between impurities 
cannot be smaller  than $R$ by definition, and there are no reasons for it to be 
substantially larger than $R$. That is why we believe that Eq.(\ref{b}) is exact at large $X$.
Our computer modeling confirms this point and it gives us the values $b_d$.

We propose an improved mean field approach which takes into account the fact that  
the density 
$\tilde{n}$ outside the circles is slightly
larger than the average density $n(R)$, because there are no impurities
inside the circles.  For example, at $d=2$, one gets
\begin{equation}
\tilde{n}=\frac{N}{S-N\pi R^2\alpha},
\label{nimpr}
\end{equation}
where $N\pi R^2\alpha$ is the excluded area inside $N$ circles.  
 We  have introduced  a free  parameter $\alpha<1$, which takes into account
the overlap of the circles.
Its value can be extracted from  comparison with numerical computations.
 
Eq. (\ref{nimpr}) can be generalized for any $d$ to get a differential equation in
$\rho_d$ 
\begin{equation}
\frac{d\rho_d}{dX^d}=-\frac{\rho_d^2}{1-\alpha_d\rho_d X^d}.
\end{equation}
The solution is given by the following  transcendental equation:
\begin{equation}
X^d=\frac{1-\rho_d^{b_d}}{b_d\rho_d}
\label{corr}
\end{equation}
with $b_d=\alpha_d+1$.
It is worth  mentioning that if we would neglect  the ``circles'' 
overlapping ($b_d\equiv 2$),
then the solution of Eq. (\ref{corr}) is
\begin{equation}
\rho_d(X)=\frac{1}{X^d+\sqrt{1+X^{2d}}},
\label{1stappr}
\end{equation}
which is an underestimate for large distances.
In the general case, for  $1< b_d<2$, 
the analytical solution of the transcendental equation (\ref{corr}) can be obtained only
for large and small values of $X$.
\begin{equation}
\rho_d\approx\left\{
\begin{array}{ll}
1-X^d +\frac{3-b_d}{2}X^{2d}+\cdots, &  X\ll 1\\
\frac{1}{b_d X^d}\left(1-\left(\frac{1}{b_dX^d}\right)^{b_d}+\cdots
\right), & X\gg 1
\end{array}\right.
\end{equation}

We performed computer simulations of this problem for the one-, two- and
three-dimensional cases. The results are shown in Fig. 2 (a) together with the simple 
mean field approximation of Eq. (\ref{n3}). 
We found that fitting  our numerical data using  Eq. (\ref{corr})
shows excellent agreement if we choose $b_1=1.67,\ b_2=1.49,\ b_3=1.15$. It would be natural 
to think that the simple  mean field approach  with $\alpha =0$ becomes exact for large 
values of $d$.

Unfortunately, Eq. (\ref{corr}) does not have an  analytical solution for  all $X$ and so 
it is not convenient for our purpose. 
We found that the simple interpolated formula
\begin{equation}
\rho_d(X)=\frac{1}{X^d+\sqrt{1+(b_d-1)^2 X^{2d}}},
\label{inrho}
\end{equation}
which resembles Eq.(\ref{1stappr}),
describes   the residual density well for the whole range of distances.
  The comparison of this formula with
the results of computer modeling is shown in Fig. 2 (b),(c),(d) for d= 1,2,3.  Below we use only Eq. (\ref {inrho})
with the values of $b_d$ obtained above.

\section{Calculation of $J(R)$}
\subsection{Three-dimensional case}
The spin-spin interaction constant is the splitting energy between
the ground states for total spin $S=1$ and $S=0$ 
$$2J=E_g^{S=1}-E_g^{S=0}\equiv ^3\Sigma_u^+ - ^1\Sigma_g^+$$
 for hydrogen-like molecule, where  nuclei are represented by two impurities.
Hereafter we use effective atomic units (a.u.) which means that all distances are
measured in units of the effective Bohr radius $a_B=\hbar^2\epsilon/m^*e^2$, and
energies in units of $m^*e^4/\hbar^2\epsilon^2$, where $m^*$ is the effective
carrier mass, and $\epsilon$ is the dielectric constant. 

We propose a simple interpolated formula for the exchange constant based on
the most accurate numerical calculations of the hydrogen molecule \cite{vol}
and the following asymptotic  expression\cite{Gor} for large $R$:
\begin{equation}
2J(R)\approx 1.636R^{5/2}\exp(-2R).
\label{as3d}
\end{equation} 

We found $J(0)$ from the data for the 
singlet-triplet splitting of the helium atom\cite{Rad}.
 $$2J(0)=0.770\ \mbox{a.u.}$$
The numerical data\cite{vol} show   that  the behavior of the logarithm of the exchange 
constant for small $R$ is well described by a second order polynomial.

To obtain the interpolated formula we match the second derivative of
$\ln\left(J(R)\right)$.
In two regions it has the following behavior
\begin{equation}\label{lnlim}
\frac{\partial^2\ln(J)}{\partial R^2}\approx\left\{
\begin{array}{ll}
-2\tilde{A}, & R\leq 1\\
-\frac{5}{2R^2}, & R\gg 1,
\end{array}\right.
\end{equation}
where $\tilde{A}$ is the matching constant. The simplest formula that satisfies
  both conditions is
\begin{equation}
\frac{\partial^2\ln(J)}{\partial R^2}=-\frac{2\tilde{A}}{1+4/5\tilde{A}R^2}
\end{equation}
 After  integrating twice we obtain
\begin{eqnarray}\label{int3D}
\ln(J) &=&\ln(J(0))-\gamma R-\frac{5}{2}AR\arctan(AR)+\frac{5}{4}\ln(1+A^2R^2),
\end{eqnarray}
where $A$ and $\gamma$ are connected by equation
\begin{equation}
A=\sqrt{4\tilde{A}/5}=\frac{4(2-\gamma)}{5\pi}.
\nonumber
\end{equation}
This interpolated formula has one fitting parameter $\gamma$ and 
the correct asymptotic behavior.

   The parameter $\gamma$ has to be chosen to match small distances in an
optimal way. The least square method gives $\gamma=0.1$. The final equation is 
\begin{equation}
\label{int3Dm}
 2J_3(R)= 0.770\left(1+0.23R^2\right)^{5/4}\exp
\left(-0.1 R-1.210 R\arctan\left(0.484 R\right)\right)\nonumber\\
\end{equation}
For further calculations we need  the inverse function
$R(J)$ as well. Because of the exponential character of the exchange
constant, the inverse function  depends on energy logarithmically.
 Therefore, we performed interpolation for the function
$R(x)$, where $x=\ln(J(0)/J)$.
The result is 
\begin{equation}\label{R3d}
R_3=\frac{x}{2}+\frac{3.5x}{1+\frac{3.5x}{1.69+0.68\ln(1+x)}}
\end{equation}  
The interpolated curves and all available data are shown in Fig. 3a.
\subsection{Two-dimensional case with in-plane impurities.}
 We are unaware of any calculations of $J(R)$  for the two-dimensional case. 
We have considered a general problem when the motion of
the electrons  is confined to a plane, but the Coulomb impurities are at distances
$h_1$ and $h_2$ outside the plane. However, in this paper  only the 
calculations for  in-plane impurities ($h_1=h_2=0$) are presented. The results for the general 
case will be  published elsewhere\cite{EFP}.

  The case of the in-plane impurities  corresponds to a $2D$ hydrogen-like 
molecule with  the   Hamiltonian 
\begin{equation}\label{H1}
\hat{H}=-\frac{\Delta_1}{2}-\frac{\Delta_2}{2}
-\sum_{j,i=1}^2\frac{1}{\sqrt{(x_i\pm a)^2 +y_i^2}} +
\frac{1}{\sqrt{(x_1-x_2)^2+(y_1-y_2)^2}}+\frac{1}{R}
\end{equation}

When $R\gg 1$ the singlet-triplet splitting constant
is calculated by making use the semiclassical approach
\cite{Gor,Smirnov,Flam99} (see Appendix \ref{ApC}). 
We obtained the following result:
\begin{equation}\label{2dH_2}
2J(R)=30.413\, R^{7/4}\exp(-4R)
\end{equation}
To provide the point $R=0$ we performed variational calculations
for the two-dimensional helium atom. We found that (See Appendix \ref{ApHel})
\begin{eqnarray}\label{2dhel}
E(^{1}S)&=&-11.635\ \mbox{a.u.}\nonumber\\
E(^{3}S)&=& -8.193\ \mbox{a.u.}\nonumber\\
2J(0)   &=&3.567\, (\pm 1\%)\  \mbox{a.u.}
\end{eqnarray}

Finally, we performed numerical calculations based on the method described in
\cite{Flam99} for the point $R=1$.
Using the same method as in the $3D$-case we get the following interpolated
formulas for $J(R)$ and $R\left(\ln(J_0/J)\right)$:

\begin{eqnarray}
 2J_2(R) &=& 3.567\left(1+1.81R^2\right)^{7/8}\exp
\left(-0.3 R-2.355R\arctan\left(1.346R\right)\right)
\label{int2D}\\
R_2  &=& \frac{x}{4}+\frac{3x}{1+\frac{3x}{0.50+0.28\ln(1+x)}},
\label{R2d}\\
x &=& \ln\left(J(0)/J)\right).\nonumber
\end{eqnarray}  

These results are shown in  Fig. 3b.

\section{Conclusion}
 We obtained an analytical expression  
(\ref{inrho}) for the dimensionless density 
of  impurities which form singlet pairs with a distance larger than $R$.
We have also calculated the strength of the spin-spin interaction $J$ and obtained  
analytical expressions  for the function $R(J)$ for the three-dimensional (\ref{R3d})
and the two-dimensional (\ref{R2d}) cases.  Combining Eq. (\ref{inrho}) with
Eqs. (\ref{R3d}) or (\ref{R2d}) one can calculate 
an analytical expression for the density of singlet pairs $n(E)/2$ that has a  
singlet-triplet energy splitting smaller 
than $E=2J$.  At finite temperature $T$ the pairs with $E<T$ are destroyed by a thermal 
motion. Therefore, at a given temperature the function $n(E)$  at $E=T$ gives the 
density of free spins in the system which contribute to the Curie susceptibility.
Thus, we obtain an analytical expression for the density of free spins $n(T)$.

We have also calculated the distribution function of excitation energy  in a 
logarithmic scale. It is defined as follows
\begin{equation}\label{Flog}
F(\ln(\varepsilon))=\frac{1}{2}\frac{d\,n}{d\,\ln(\varepsilon)}
\equiv \frac{n_0}{2}\frac{d\,\rho}{d\,R}\frac{d\,R}{d\,\ln(\varepsilon)},
\end{equation}
where $\varepsilon =E/(2J(0))\equiv J/J(0)$.

The analytical expression for $F(\ln(\varepsilon))$ based on Eqs. (\ref{inrho}),
(\ref{R3d}), (\ref{R2d}) is quite cumbersome.
In the two limits of large and small
energies (or small and large distances) the behavior of $F$ in the
leading order is
\begin{equation}\label{F2D}
\frac{F_{2D}}{n_0}\approx\left\{\begin{array}{ll}
10.56\pi n_0 a_B^2\ln(1/\varepsilon), &\qquad 
\varepsilon\longrightarrow 1\\
\frac{16}{1.49\pi n_0 a_B^2}\left[\ln(1/\varepsilon)\right]^{-3}, &\qquad \varepsilon
\longrightarrow 0;
\end{array}\right.
\end{equation}
\begin{equation}\label{F3D}
\frac{F_{3D}}{n_0}\approx\left\{\begin{array}{ll}
 128\pi n_0 a_B^3\left[\ln(1/\varepsilon)\right]^2, &\qquad 
\varepsilon\longrightarrow 1\\
\frac{9}{1.15\pi n_0 a_B^3}\left[\ln(1/\varepsilon)\right]^{-4}, &\qquad \varepsilon
\longrightarrow 0.
\end{array}\right.
\end{equation}

Our results for  $\rho (T)=n(T)/n_0$ are shown by the full lines in Fig. 4 (a), (b) 
for the two-dimensional and the three-dimensional cases. We choose two different
dimensionless densities $n_0$ for each
case. They are $\pi n_0 a_B^2=0.1$ and 0.025 for $2D$ and $4\pi n_0 a_B^3/3=0.004$ and 0.016 
for $3D$.

The dependence of  the dimensionless distribution function $F/n_0$ for the two-dimensional and the
three-dimensional cases for the same two donor densities $n_0$  are presented on Fig. 4
(c), (d) by the full lines.

The most important features of both functions are the long logarithmic tails in the regions 
of low temperature and low energy. Similar behavior has been obtained by Bhatt and Lee\cite{lee}. 
Note that  $\rho(T)$ decreases with increasing density  $n_0$. This is not the case for the
distribution function. Larger density corresponds to larger distribution function $F$ at large
energies. This is because the derivative $n_0^{-1} dn/dR$ is larger for larger density $n_0$ at
small $R$. However,  the dimensionless distribution function $F(\ln \varepsilon )$
 is normalized to 1/2. That is why  the functions for different densities
 cross each other at some energy.
 Thus, at
small energies the larger distribution function corresponds to smaller density.

To clarify the role of the functions $J(R)$,
which have been found here, we have calculated the density of free spins  $\rho (T)$ and the
distribution function
$F(\ln(\varepsilon))$  for a simplified function $J_s(R)$ used in Ref.\cite{lee}.
\begin{eqnarray}\label{epslee}
J_s(R)&=& J(0)\exp(-2R/a_B) \qquad \mbox{for 3D,}\nonumber\\
J_s(R)&=& J(0)\exp(-4R/a_B) \qquad \mbox{for 2D.}
\end{eqnarray}
In these calculations we used  our  residual density $\rho(R)$. The results are shown 
in Fig. 4 by the dashed lines.  One can see that the difference is large.  The
distribution function for our more accurate form of the exchange constant became narrower and 
the decline  in the beginning is steeper. This   comes from the different behaviors
at small distances.  

It is interesting to compare the numerical scaling calculations by Bhatt and Lee\cite{lee}
with our method of calculation $n(R)$. We have found that at the smallest density used 
in Ref. \cite{lee} both methods give 
similar results, but for larger densities there is a small deviation. We think that both 
methods are exact in the limit of small densities, but the method of Bhatt and Lee 
works in a wider range, because they take into
account the renormalization of weak bonds caused by their strong neighbors.
However,  the great advantage of our method 
is that it gives an analytical expression for $n(E)$.

\acknowledgments
This work is supported by the Australian Research
Council and by the Seed Grant of the University of Utah.  A. L. Efros and V. V. Flambaum  are 
grateful to the Center for Theoretical Physics in Trieste for hospitality during
 the work on this
project. A. L. Efros is grateful to R. Bhatt, D. Mattis, and B. Sutherland for helpful 
discussions. I. Ponomarev acknowledge fruitful discussions with M. Kuchiev and G. Gribakin. 
\appendix

\section{Exchange constant for $2D$ Hydrogen-like molecule}\label{ApC}
The exchange constant for the Hamiltonian (\ref{H1}) is determined by
\begin{eqnarray}
2J &=&(E_g^{S=1}-E_g^{S=0})\  \equiv (E_A-E_S),\quad
\mbox{ where }\label{J}\\
\hat{H}\Psi_S &=& E_S\Psi_S,\label{eq1}\\
\hat{H}\Psi_A &=& E_A\Psi_A.\label{eq2}
\end{eqnarray}
Because of the Fermi statistics the two-electron wave function
 is antisymmetric with
respect to permutation. Therefore, the symmetric coordinate wave function
corresponds to spin $S=0$ and the antisymmetric one corresponds to $S=1$.

Let us consider the more general Hamiltonian 
\begin{equation}\label{H12d}
\hat{H}=-\frac{\Delta_1}{2}-\frac{\Delta_2}{2}
+V_a({\bf r}_1+{\bf a})+V_a({\bf r}_2+{\bf a})
+V_b({\bf r}_1-{\bf a})+V_b({\bf r}_2-{\bf a})
+\frac{1}{|{\bf r}_1-{\bf r}_2|}+\frac{1}{R},
\end{equation}
 where $V_a$ and $V_b$ are effective potentials of interaction between
the electron and the corresponding atomic residue, which is of the Coulomb type
 far from the atoms: $V_{a,b}\rightarrow -1/r,\ r-\rightarrow\infty$.
The electron energy
$$E=-\alpha^2/2-\beta^2/2-1/R,$$
is accurate up to terms $\sim 1/R^2$. Here
$\alpha^2/2$ and $\beta^2/2$ are electron binding energies in the given
``atom''.

When $R\gg 1$ the most appropriate method for determination of  
the energy terms splitting due to the spin-spin interaction
is the Gor'kov-Pitaevskii method \cite{Gor,Smirnov,Flam99}.

 Since $J(R)$ is exponentially small as $R\rightarrow\infty$,
$\Psi_A$ and $\Psi_S$ are solutions of the same Schr\"{o}dinger equation,
and therefore, with exponential accuracy their combinations
$$\Psi_{1,2}=\frac{\Psi_S\pm\Psi_a}{\sqrt{2}}$$ are also the solutions
of the same Schr\"{o}dinger equation with  the Hamiltonian (\ref{H12d}).
 They correspond to the states of ``distinguishable''
particles, when, e.g. for $\Psi_1({\bf r_1, r_2})$, the first electron
is principally located near the first ion at $x=-a$
and the second electron near the second ion with $x=a$. 
Here $R\equiv 2a$ is the distance between ``nuclei'',
which we place at the points $\pm a$ on the $x$ axis.
In the main region of the electron distribution, the wave functions $\Psi_{1,2}$
are products of the atomic single-particle wave functions with the asymptotic behavior
of the radial atomic wave functions of the electron in the Coulomb field of the
atomic residue being determined by the formulas
\begin{equation}\label{assymH}
\varphi_a(r)=A_{\alpha}r^{1/\alpha-1/2}e^{-\alpha r},\qquad 
\varphi_b(r)=A_{\beta}r^{1/\beta-1/2}e^{-\beta r}. 
\end{equation}
Indeed, for large $r$ the potential is $U\sim -1/r$, and the single-particle wave 
function of the electron obeys the equation
$$-\frac{\Delta}{2}\varphi-\frac{1}{r}\varphi=-\frac{\alpha^2}{2}\varphi.$$
It has the asymptotic solution (\ref{assymH})
up to  $\varphi/r^2$ accuracy. The coefficients $A_{\alpha,\beta}$ are determined
by the behavior of the wave functions of the electron inside the atoms.
 
It is possible to show\cite{Flam99}  that 
\begin{equation}\label{JvsInt}
2J=-2\int\left [\Psi_{2}\frac{\partial\Psi_1}{\partial x_1}-
\Psi_{1}\frac{\partial\Psi_2}{\partial x_1}
\right]_{x_1=x_2}~dx_2~dy_1~dy_2.
\end{equation}

 Our main purpose is to find the wave function $\Psi_{1,2}$.

Let us suppose that
$\Psi_{1,2}$ has the form
\begin{eqnarray}
\Psi_1(\vec{r}_1,\vec{r}_2)&=&\phi_{\alpha}(|\vec{r}_1+\vec{a}|)
\phi_{\beta}(|\vec{r}_2-\vec{a}|)\chi(\vec{r}_1,\vec{r}_2),\nonumber\\
\Psi_2(\vec{r}_1,\vec{r}_2)&=&\phi_{\alpha}(|\vec{r}_2+\vec{a}|)
\phi_{\beta}(|\vec{r}_1-\vec{a}|)\chi(\vec{r}_2,\vec{r}_1),
\end{eqnarray}
where $\phi_{\alpha,\beta}$ have the behavior of (\ref{assymH}) and
$\chi$ is a slowly varying function of $r_1$ and $r_2$.
 Substituting $\Psi_1$ into the wave equation and neglecting the second derivatives
of $\chi$, we obtain
\begin{equation}\label{chieq}
\alpha \frac{\partial\chi}{\partial x_1} -
\beta \frac{\partial \chi}{\partial x_2}
+\left [\frac{1}{\sqrt{(x_1-x_2)^2+(y_1-y_2)^2}}+\frac{1}{2a}
-\frac{1}{a-x_1}-\frac{1}{a+x_2}\right ]\chi =0.
\end{equation}
Equation (\ref{chieq}) is valid under the conditions
\begin{eqnarray}\label{chicon}
&& |x_{1,2}|\leq a, y_{12}\equiv |y_1-y_2|\ll\sqrt{a}\nonumber\\
&& R\alpha,R\beta\gg 1, R|\alpha-\beta| \ll 1.
\end{eqnarray}
 The general solution of (\ref{chieq}) is
\begin{equation}\label{gens}
F(C_1(x_1,x_2),C_2(\chi,x_1,x_2,y_{12})=0,
\end{equation}
where $C_1,\  C_2$ are integrals of the motion of the ordinary
differential equations:
$$ \frac{dx_1}{\alpha}=-\frac{dx_2}{\beta}=-\frac{d\chi}{\chi}
\left [\frac{1}{\sqrt{(x_1-x_2)^2+y_{12}^2}}+\frac{1}{2a}
-\frac{1}{a-x_1}-\frac{1}{a+x_2}\right ]^{-1}.
$$
 Hence
\begin{equation}\label{chisol}
\chi(x_1,x_2,y_{12})=\frac{\exp\left (-\frac{x_1}{2a\alpha}\right )
\left [\sqrt{(x_1-x_2)^2+y_{12}^2}-x_1+x_2\right ]^{\frac{1}{\alpha+\beta}}
}{[a-x_1]^{1/\alpha}[a+x_2]^{1/\beta}}f\left (\frac{x_1}{\alpha}+
\frac{x_2}{\beta}\right ),
\end{equation}
where the unknown function $f(u)$ is determined from
the fact that  $\chi\longrightarrow 1$ when $x_1\longrightarrow -a,\ x_2$
is arbitrary, or when $x_2\longrightarrow a$ and $x_1$ is arbitrary.
  Finally, after expanding $|\vec{r}\pm a|\simeq |a\pm x|+y_{12}^2/2|a\pm x|$
in the exponent, we obtain
\begin{eqnarray}\label{wffin}
\Psi_1(\vec{r}_1,\vec{r}_2)&=&A_{\alpha}A_{\beta}\left (a+x_1\right )^
{\frac{2-\alpha}{2\alpha}}\left (a-x_2\right )^
{\frac{2-\beta}{2\beta}}\times\nonumber\\
& &\times \exp\left [-a(\alpha+\beta)+\beta x_2-
\alpha x_1-\frac{\alpha y_1^2}{2(a+x_1)}-\frac{\beta y_2^2}{2(a-x_2)}
\right ]\chi(\vec{r}_1,\vec{r}_2),\\
\chi(x_1,x_2,y_{12})&=&\nonumber
\end{eqnarray}
\begin{equation}
\left\{
\begin{array}{l}
e^{-\frac{a+x_{1}}{2a\alpha}}
\left[\frac{2a}{a-x_1}\right]^{\frac{1}{\alpha}}
\alpha^{-\frac{\alpha}{(\alpha+\beta)\beta}}
\left[\frac{\beta (a+x_1)+\alpha (a+x_2)}{a+x_2}\right]^{\frac{1}{\beta}}
\left[\frac{\sqrt{(x_1-x_2)^2+y_{12}^2}-x_1+x_2}
{\sqrt{\left(\beta (a+x_1)+\alpha (a+x_2)\right)^2+(\alpha y_{12})^2}
+\beta (a+x_1)+\alpha (a+x_2)}\right]^{\frac{1}{\alpha+\beta}}
 \\
e^{-\frac{a-x_{2}}{2a\beta}}
\left[\frac{2a}{a+x_2}\right]^{\frac{1}{\beta}}
\beta^{-\frac{\beta}{(\alpha+\beta)\alpha}}
\left[\frac{\beta (a-x_1)+\alpha (a-x_2)}{a-x_1}\right]^{\frac{1}{\alpha}}
\left[\frac{\sqrt{(x_1-x_2)^2+y_{12}^2}-x_1+x_2}
{\sqrt{\left(\beta (a-x_1)+\alpha (a-x_2)\right)^2+(\beta y_{12})^2}
+\beta (a-x_1)+\alpha (a-x_2)}\right]^{\frac{1}{\alpha+\beta}}
\end{array}\right.
\end{equation}
Here the upper expression is given for $x_1+x_2\leq 0$, and the lower expression
for $x_1+x_2\geq 0$.

Substituting (\ref{wffin}) in Eq. (\ref{JvsInt}), and differentiating
only the exponential we obtain
\begin{equation}\label{Jgen}
2J=+2(\alpha+\beta)\int_{-a}^a\left[\Psi_1\Psi_2\right]_{x_1=x_2=x}\,
d\,x\,d\,y_1d\,y_2
\end{equation}
Introducing the notations $\mu=\alpha+\beta$ and $\nu=\beta-\alpha$ and
taking into consideration the fact that at the approximation (\ref{chicon})
$$ \sqrt{\left[(\beta+\alpha)(a-x)\right]^2+(\beta y_{12})^2}
+(\beta+\alpha)(a-x)\approx 2(\beta+\alpha)(a-x).$$
the formula (\ref{Jgen}) transforms to 

\begin{equation}\label{Jalbet}
2J(\alpha,\beta,R)=
R^{\frac{2}{\alpha}+\frac{2}{\beta}-\frac{1}{\mu}}e^{-\mu R}
\left[D\left(\alpha,\beta,R\right)+D\left(\beta,\alpha,R\right)\right],
\end{equation}

where $D\left(\alpha,\beta,R\right)$ is the  following function:
\begin{eqnarray}\label{Dabr}
D\left(\alpha,\beta,R\right)&=&
4\sqrt{\pi}A_{\alpha}^2 A_{\beta}^2
\left(\frac{\mu}{2}\right)^{-1/\mu}
\Gamma\left(\frac{2+\mu}{2\mu}\right)
\left(2^{-\mu/\alpha}\frac{\mu}{\alpha}\right)^{
\frac{2\alpha}{\mu\beta}}\times\nonumber\\
& &\int_0^1\frac{
\exp\left(-(1-x)/\alpha-\nu Rx\right)
(1+x)^{2/\beta-2/\alpha+1/\mu}
(1-x)^{2/\alpha-1/\mu}
}{\left(1-x\nu/\mu\right)^{1+1/\mu}} dx.
\end{eqnarray}
In the case $\alpha=\beta$ it is independent of $R$:
\begin{equation}\label{D0a}
D_0(\alpha)\equiv 2\,D\left(\alpha,\alpha,R\right)=
8\sqrt{\pi}A_{\alpha}^4\left(\frac{1}{4\alpha}\right)^{1/2\alpha}
\Gamma\left(\frac{\alpha+1}{2\alpha}\right)
\int_0^1 \exp\left(-t/\alpha\right)
(2-t)^{1/2\alpha}t^{3/2\alpha}\,dx,
\end{equation}
and
\begin{equation}
\label{Jf} 
2J(\alpha,\alpha,R)=D_0(\alpha)R^{7/2\alpha}\exp(-2\alpha R)
\end{equation}

For the two-dimensional hydrogen molecule ($\alpha=2,\ 
A_{\alpha}=4/\sqrt{2\pi}$) it gives
\begin{equation}\label{H_2}
J(2,2,R)=30.413\, R^{7/4}\exp(-4R)
\end{equation}

\section{Two dimensional helium atom}\label{ApHel}
\subsection{Variational method}

 To find $J(0)$ one should consider the singlet-triplet splitting of two impurities which are at a
distance much smaller than the Bohr radius of one impurity state. The motion of electrons is
restricted by the plain, so  this is as a ``two-dimensional helium atom''.
 In this case the variational approach is the most appropriate.
The Hamiltonian is:
\begin{equation}\label{HHe}
\hat{H}=-\frac{\Delta_1}{2}-\frac{\Delta_2}{2}
-\frac{Z}{r_1} -\frac{Z}{r_2}+\frac{1}{r_{12}},
\end{equation}
and Schr\"{o}dinger's variational principle is
\begin{eqnarray}\label{vsp}
E&=& \min \frac{\int \Psi\hat{H}\Psi\,d\tau}{N},\qquad \mbox{ where}\nonumber\\
N&=& \int \Psi^2\,d\tau
\end{eqnarray}
The most important thing 
is the correct choice of the coordinate system. Namely, it is better to choose as
independent variables those that the potential energy depends on.
These are the three sides of the triangle $r_1$, $r_2$, $r_{12}$ between
the nucleus and two electrons\cite{Bethe}.
  The Hamiltonian and, as we expect, the wave functions for  $S$ terms 
do not  depend on the orientation of the triangle in the space:
\begin{equation}\label{vwfl}
\Psi(\vec{r}_1,\vec{r}_2)=\Psi(r_1,r_2,\theta)=\Psi(r_1,r_2,-\theta) 
\end{equation}
\begin{equation}
r_{12}^2=r_1^2+r_2^2-2r_1r_2\cos(\theta)
\end{equation}
Therefore, the volume element $d\tau$ is
$$d\tau=r_1r_2\,dr_1\,dr_2\,d\phi\,d\theta=8\pi \frac{r_1r_2r_{12}
\,dr_1\,dr_2\,dr_{12}}{\sqrt{4r_1^2r_2^2-\left(r_1^2+r_2^2-r_{12}^2\right)^2}}
$$
 Finally, we introduce the ``elliptic'' coordinates 
\begin{eqnarray}\label{elcoor}
s&=&r_1+r_2,\nonumber\\
t&=&r_1-r_2,\nonumber\\
u&=&r_{12},
\end{eqnarray}
which reflect the symmetry of two-particle eigenfunction:
the wave function  has to be an even function of $t$ for total spin $S=0$,
and  an odd function of $t$ for $S=1$. 
Thus,
\begin{equation}
d\tau=\frac{\pi\left(s^2-t^2\right)u}{\sqrt{(s^2-u^2)(u^2-t^2)}}\,ds\,dt\,du
\end{equation}
The factor $\pi$ can be omitted, and if we take into consideration the fact that
$$\Psi^2(s,t,u)=\Psi^2(s,-t,u),\quad \Psi\hat{H}\Psi(s,t,u)=
\Psi\hat{H}\Psi(s,-t,u),$$
 we can restrict the integration region by the inequalities:
\begin{eqnarray}
&&0\leq t\leq u\leq s\leq\infty \\
\mbox{ (or }&& t\leq u\leq s,\ 0\leq t\leq s \leq \infty)
\end{eqnarray}
The potential energy in the new coordinates is
\begin{equation}\label{poten}
\langle\Psi|-\frac{Z}{r_1} -\frac{Z}{r_2} +
\frac{1}{r_{12}}|\Psi\rangle=
\langle\Psi|-\frac{4Zs}{s^2-t^2}|\Psi\rangle+\langle\Psi|\frac{1}{u}
|\Psi\rangle=V_1+V_2
\end{equation}
And the mean value of the kinetic energy is
\begin{eqnarray}\label{kinen}
K&=& \langle \Psi|-\frac{\Delta_1}{2}-\frac{\Delta_2}{2}|\Psi\rangle=
\frac{1}{2}\int\,\left[
\left(\nabla_1\Psi\right)^2+\left(\nabla_2\Psi\right)^2\right]\,d\tau=
\nonumber\\
&=&\int_0^{\infty}ds\int_0^s du\int_0^u\left(
\frac{u(s^2-t^2)}{\sqrt{(s^2-u^2)(u^2-t^2)}}\left[
\left(\frac{\partial\Psi}{\partial s}\right)^2+
\left(\frac{\partial\Psi}{\partial t}\right)^2+
\left(\frac{\partial\Psi}{\partial u}\right)^2\right]\right.+\nonumber\\
&+&\left. 2\frac{\partial\Psi}{\partial u}\left[
s\sqrt{\frac{u^2-t^2}{s^2-u^2}}\frac{\partial\Psi}{\partial s}+
t\sqrt{\frac{s^2-u^2}{u^2-t^2}}\frac{\partial\Psi}{\partial t}
\right]\right)dt 
\end{eqnarray}  
\subsection{Ground state of He}
For the ground state we use the trial wave function in a form
\begin{eqnarray}\label{grwf}
\Psi&=&\frac{1}{2}\left(e^{-(\alpha_1r_1+\alpha_2r_2)}+ 
e^{-(\alpha_2r_1+\alpha_1r_2)}\right)=e^{-\alpha s}\cosh(\beta t)\nonumber\\
\alpha&=&(\alpha_1+\alpha_2)/2\nonumber\\
\beta &=&(\alpha_1-\alpha_2)/2
\end{eqnarray}
and we also introduce the parameter $\gamma=(\beta/\alpha)^2$.

After calculating all necessary integrals we obtain:
\begin{eqnarray}\label{Nor}
N^{-1}&=&\frac{(2\alpha)^4}{\pi}\frac{(1-\gamma)^2}{1+(1-\gamma)^2}\nonumber\\
K/N&=&\alpha^2\frac{1+\gamma+(1-\gamma)^3}{1+(1-\gamma)^2}\nonumber\\
V_1/N&=&=-4\alpha Z\nonumber\\
V_2/N&=&=\frac{\alpha}{2}\frac{(1-\gamma)^2}{1+(1-\gamma)^2}\left(
\frac{3\pi}{4}+F(\gamma)\right),
\end{eqnarray}
\begin{equation}
F(\gamma)=\int_0^1\frac{2-u^2+
\gamma u^2(1-2u^2)}{\sqrt{1-u^2}(1-\gamma u^2)^{5/2}}du,
\end{equation}

where $N,V_1,V_2,$ and $K$ are defined by Eqs. (\ref{vsp}),(\ref{poten}), and
(\ref{kinen}) correspondingly. 
Here, in order to find $V_2$ we used the values of the following integrals:
\begin{eqnarray}\label{intv21}
\int_0^1\cosh(bt)\left[1-t^2\right]^{\mp 1/2}dt&=&\frac{\pi}{2}\left\{  
\begin{array}{ll}
I_0\left(b\right) &\qquad \mbox{(see \cite{Rizhik}, 3.534.2)} \\
I_1\left(b\right)/b, & 
\end{array}\right.
\nonumber\\
\int_0^{\infty}x^{n-1}e^{-x}\frac{I_{\nu}(cx)}{c^{\nu}}dx
&=&
(-1)^{n-1}\frac{\partial^{n-1}}{\partial p^{n-1}}\left[\frac{
(p+\sqrt{p^2-c^2})^{-\nu}}{\sqrt{p^2-c^2}}\right]_{p=1}
\quad \mbox{ (see \cite{prud}, 2.15.3)}\nonumber
\end{eqnarray}

Thus, the energy is
\begin{equation}
E(\alpha,\gamma)=\alpha^2\frac{1+\gamma+(1-\gamma)^3}{1+(1-\gamma)^2}
-\alpha\left(4Z-\frac{1}{2}\frac{(1-\gamma)^2}{1+(1-\gamma)^2}\left(
\frac{3\pi}{4}+F(\gamma)\right)\right)
\end{equation}
We can also rewrite it in the form:
\begin{equation}
E(\alpha,\gamma)=\alpha^2 f(\gamma)-\alpha g(\gamma)
\end{equation}
The minimum value of the energy is realized for values of $\alpha$ and
$\gamma$ which satisfy the equations
$$
\frac{\partial E}{\partial \alpha}=0,\qquad 
\frac{\partial E}{\partial \gamma}=0,
$$
or
$$
2\alpha f(\gamma)-g(\gamma)=0, \alpha f'(\gamma)-g'(\gamma)=0.
$$
Eliminating $\alpha$, we get the equation in $\gamma$
\begin{equation}
\frac{f'(\gamma)}{f(\gamma)}-2\frac{g'(\gamma)}{g(\gamma)}=0.
\end{equation}
It has the solution $\gamma=0.11436$.

Then,
\begin{eqnarray}\label{E2par}
\alpha &=& 3.4059,\nonumber\\
E_S &=& -11.760.
\end{eqnarray}
 The corresponding values of $\alpha_1$ and $\alpha_2$ are
\begin{eqnarray}
\alpha_1&=&\alpha (1+\sqrt{\gamma})=4.5576\nonumber\\
\alpha_2&=&\alpha (1-\sqrt{\gamma})=2.2541
\end{eqnarray}
 For comparison we also represent the results for the one-parameter wave
function $\Psi=e^{-\alpha_1(r_1+r_2)}=e^{-\alpha s}$:
\begin{eqnarray}\label{E1par}
E(\alpha)&=&\alpha^2-\alpha(8-\frac{3\pi}{8})\nonumber\\
\frac{\partial E}{\partial \alpha}=0 &\Rightarrow& \alpha=4-\frac{3\pi}{16}
=3.4109\\
E&=& -11.635.
\end{eqnarray}
 Thus, the difference between these two  ground state energies
(\ref{E2par}) and (\ref{E1par}) is $1\%$. 
\subsection{Term $^3S$ of He}
Taking into consideration the screening effect of the electrons we
construct our trial wave function from an antisymmetric combination
of the $1s$-electron in a field of the charge $Z_1\equiv\alpha_1/2$ and
the orthogonal $2s$-electron state in a field of the charge 
$Z_2\equiv 3\alpha_2/2$: 
\begin{eqnarray}
\varphi_{10}(r)&=&\exp\left(-\alpha_1 r\right),\nonumber\\
\varphi_{20}(r)&=&\exp\left(-\alpha_2 r\right)\left(1-
\frac{\alpha_1+\alpha_2}{2}r\right),\\
\Psi(r_1,r_2)&=&\varphi_{10}(r_1)\varphi_{20}(r_2)-
\varphi_{10}(r_2)\varphi_{20}(r_1).
\end{eqnarray}
Or rewriting in $s$, $t$ variables:
\begin{eqnarray}
\Psi(s,t)&=&e^{-\alpha s}\left[\alpha t\cosh(\beta t)+
\left(\alpha s-2\right)\sinh(\beta t)\right],\\
\alpha &=&\frac{\alpha_1+\alpha_2}{2},\nonumber\\
\beta&=&\frac{\alpha_1-\alpha_2}{2}. 
\end{eqnarray}
Performing a procedure  similar to, though more tedious than, the $^1S$ case 
we obtain the following results ($\lambda=\beta/\alpha$):
\begin{eqnarray}
E(\alpha,\lambda)&=&\alpha^2\frac{K}{N}-\alpha\frac{V_1+V_2}{N}\label{E3_s}\\ 
N &=& \frac{2\pi}{(2\alpha)^4}\frac{2\lambda^2+1}{(1-\lambda)^4(1+\lambda)^2},
\nonumber\\
K/N&=&2\frac{\lambda^4-\lambda^3+4\lambda^2-2\lambda+1}{2\lambda^2+1},
\nonumber\\
V_1/N &=&8\frac{3\lambda^2-\lambda+1}{2\lambda^2+1},\nonumber\\
V_2/N &=&-\frac{1}{16}\frac{(1-\lambda)^4(1+\lambda)^2
\left(F(\lambda)-171\pi/16\right)}{2\lambda^2+1}.
\end{eqnarray}
Here 
\begin{eqnarray}
F(\lambda)&=&
\int_0^1\frac{\phi(u,\lambda)}{\left (1-{\lambda}^{2}{u}^{2}\right )^{9/2}
\sqrt {1-{u}^{2}}}\,du,\nonumber\\
\varphi(u,\lambda)&= &16\, u^6 \left(1-2\,u^2\right )\lambda^6+
24\,u^6\left (1-2\,u^2\right ){\lambda}^{5}+\nonumber\\
&+& \left (45\,{u}^{4}-24\,{u}^{8}\right ){\lambda}^{4}+
\left (162\,{u}^{4}-144\,{u}^{6}\right )\lambda^3+
 12\,u^2\left (1+3\,u^2-6\,u^4 \right ){\lambda}^{2}+\nonumber\\
&+&
6\,u^2\left (4-3\,u^2\right )\lambda +
32-4\,u^2-9\,u^4
\end{eqnarray}
From minimizing $E(\alpha,\lambda)$ we find
\begin{equation}
\lambda=0.74217, \alpha=2.30998,
\end{equation}
which correspond to the effective charges $Z_1=2.0122$ and $Z_2=0.8934$.
For these values of $\lambda$ and $\alpha$ we get the energy
\begin{equation}
E_A=-8.19345
\end{equation}
It is worthwhile to note that for the wave function with $Z_1=2$ and
$Z_2=1$ the energy $E=-8.19062$ is only higher by $0.03\%$!

 Thus the exchange constant for the $2D$ Helium atom is
\begin{equation}\label{JHe}
J=E_A-E_S=3.567\  (\pm 1\%) \mbox{ a.e.}
\end{equation}
 
\begin{figure}\label{fig1}
\psfig{file=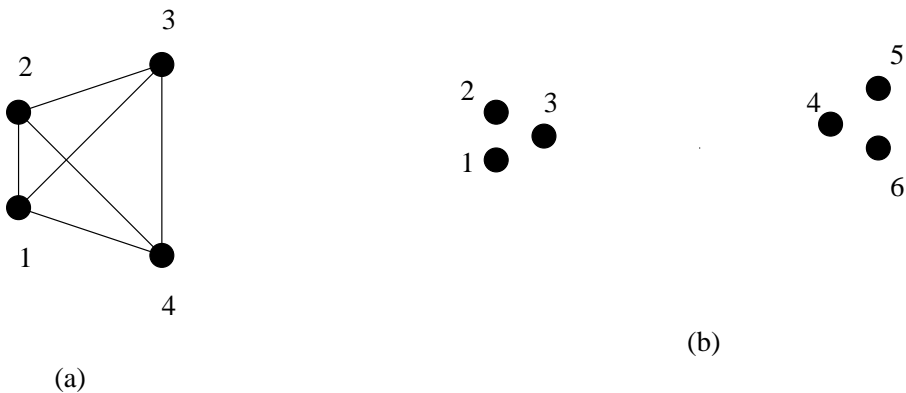,angle=270}
\caption{Different configurations of 4 (a) and 6 (b) spins.}
\end{figure}
\begin{figure}\label{fig2}
\psfig{file=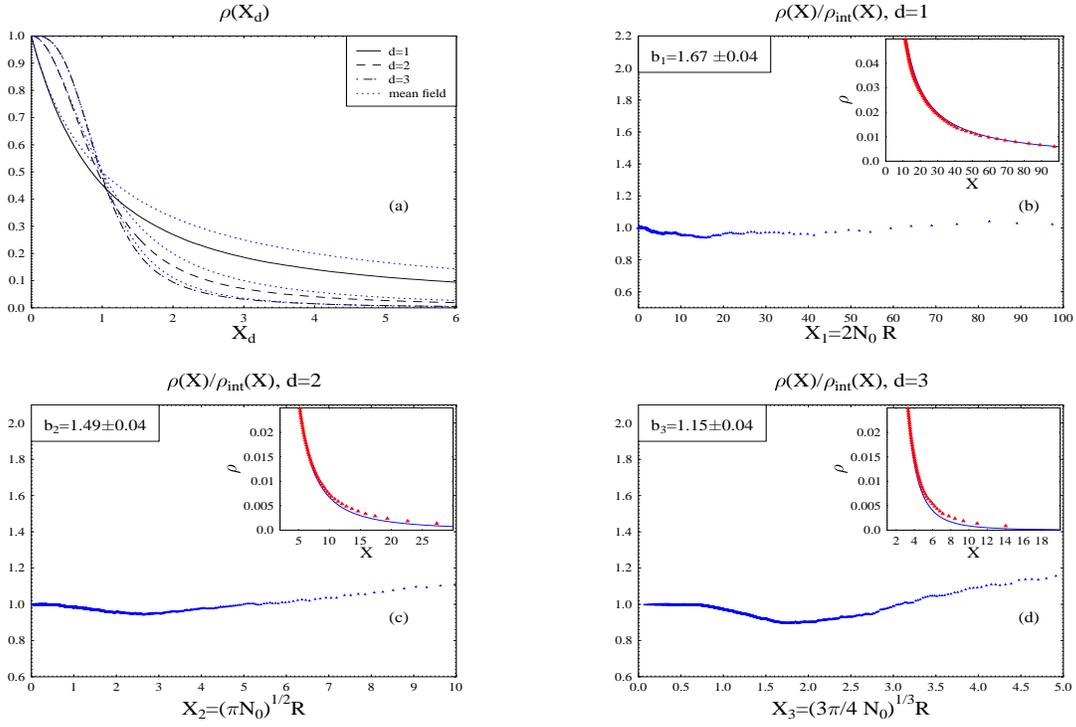,height=14cm,width=12cm,angle=270}
\caption{Residual density of impurities as a function of $X$.  Fig. 2(a)
shows the results of numerical computations for d=1,2,3 and the result of
a simple  mean field approximation  Eq. (\protect\ref{n3}).  Fig. 2 (b)-(d) 
displays the ratio of $\rho_d$ as obtained by numerical computations to 
$\rho_d$ found using the interpolated formula Eq. (\protect\ref{inrho}) for 
$d$=1-3. The insets show the asymptotic behavior of the numerical data (triangles) and the
interpolated formula (solid line).}
\end{figure}
\begin{figure}\label{fig3}
\psfig{file=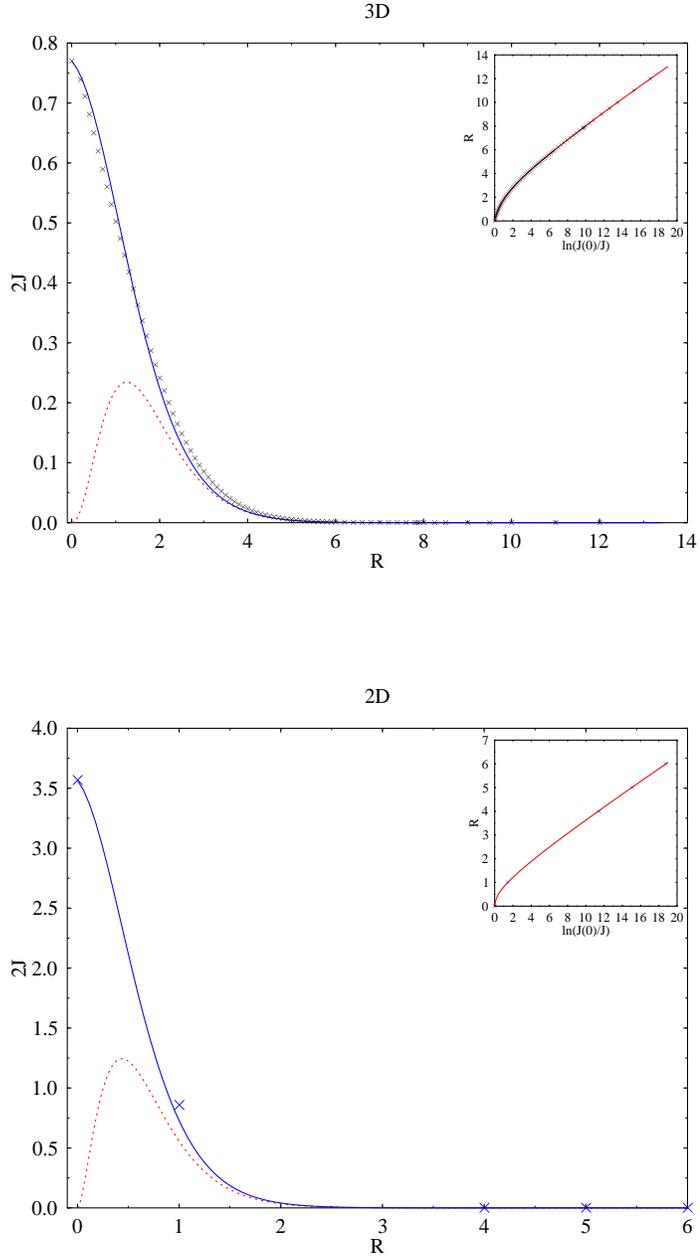,height=19cm}
\caption{Interpolated formulae (solid lines) for $J(R)$ in the three- and two 
dimensional cases as given by Eqs. (\protect\ref{int3Dm},\ref{int2D}). 
For  $3D$-case the crosses show numerical results  of 
\protect\cite{vol}. For $2D$-case  the crosses show  our  numerical results at
$R=1$ and and $R=0$.
Dashed lines show  
the corresponding asymptotic formulae which are valid at large $R$. The solid lines in the insets
show the behavior of the inverse function $R$ vs. $\ln(J(0)/J)$ as given by interpolated formulae 
(\protect\ref{R3d}) and
(\protect\ref{R2d}). The crosses in the insets have the same meaning as in the main figures.}
\end{figure}
\begin{figure}\label{fig4}
\psfig{file=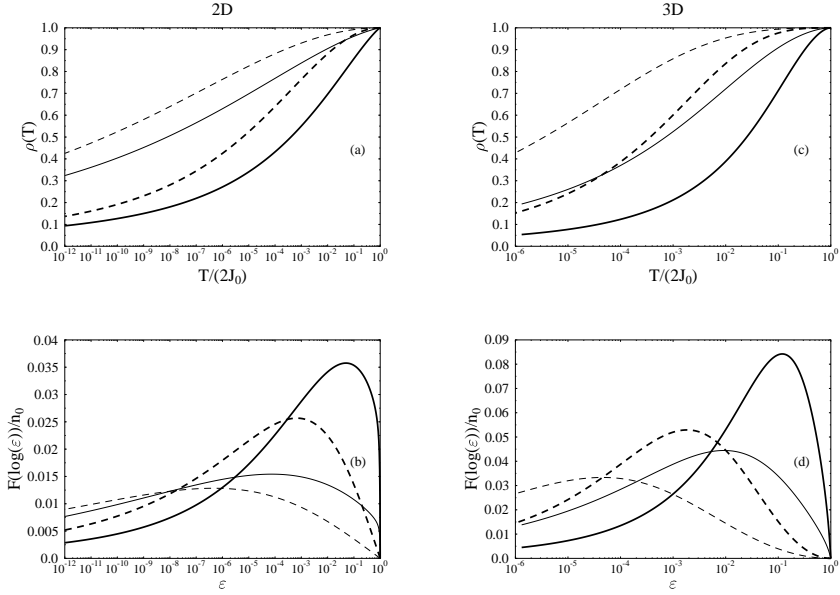,height=12.5cm,angle=270}
\caption{Dimensionless  density of free spins $\rho(T)$ as a function of logarithm of temperature
and dimensionless distribution function $F/n_0$ of the singlet-triplet excitations as a function 
of logarithm of  energy in two-dimensional case (a),(b) and three-dimensional case (c), (d). The
values of dimensionless density $\pi n_0 a_B^2$ at (a) and (b) are 0.025 and 0.1. The values of
dimensionless density $4\pi n_0 a_B^3/3$ at (c) and (d) are 0.004 and 0.016. The solid lines show
the final results of our calculations. The dashed lines show the results obtained with the function
$J(R)$ as given by Eqs. (\ref{epslee}). In all cases larger densities are shown by a thicker 
lines.}
\end{figure}

\end{document}